\documentstyle[twocolumn,aps]{revtex}

\begin{document}
\draft

\title{
Edge states in Open Antiferromagnetic Heisenberg Chains
}
\author{Shaojin Qin}
\address{
Institute of Theoretical Physics,
P. O. Box 2735,
Beijing 100080, P.R. China
}
\author{Tai-Kai Ng}
\address{
Department of Physics,
Hong Kong University of Science and Technology,\\
Clear Water Bay Road,
Kowloon, Hong Kong
}
\author{Zhao-Bin Su}
\address{
Institute of Theoretical Physics,
P. O. Box 2735,
Beijing 100080, P.R. China
}
\date{Dec 13, 1994}
\maketitle
\begin{abstract}
  In this letter we report our results in investigating edge effects of
open antiferromagnetic Heisenberg spin chains with spin magnitudes $S=1/2,
1,3/2,2$ using the density-matrix renormalization group (DMRG) method
initiated by White. For integer spin chains, we find that edge states with
spin magnitude $S_{edge}=S/2$ exist, in agreement with Valence-Bond-Solid
model picture. For half-integer spin chains, we find that no edge states
exist for $S=1/2$ spin chain, but edge state exists in $S=3/2$ spin chain
with $S_{edge}=1/2$, in agreement with previous conjecture by Ng.  Strong
finite size effects associated with spin dimmerization in half-integer
spin chains will also be discussed.
\end{abstract}


\narrowtext

The antiferromagnetic Heisenberg spin chains has been a subject of immense
interest in the last decade since Haldane conjectured that the low
energy physics of integer and half integer spin chains are fundamentally
different\cite{hal}. It is now generally believed that half-integer Heisenberg
spin chains have gapless excitation spectrum, whereas gaps exist in integer
spin chains (Haldane gap). More recently, there has been increasing interests
in studies of spin chains with defects\cite{e1,e2,e3,ea}. In particular, the
properties of broken $S=1$ quantum spin chains have received much attention
because of the experimental observation of $S=1/2$ excitations localized
at the ends of broken $S=1$ spin chains\cite{e1}. More generally, one
may address the question of whether edge states are genuine properties of
finite quantum spin chains as in Fractional Quantum Hall Effect\cite{wen,mac}.
Recently, a theory of edge states based on the Non-linear-sigma model ($NL
\sigma{M}$) plus topological $\theta$-term has been developed by one
of us\cite{ng} where it was conjectured that edge states are genuine
properties of antiferromagnetic quantum spin chains with spin value
$S>1/2$. In this letter, we shall address this question for both integer
and half integer spin chains numerically using the recently developed density
matrix renormalization group (DMRG) method\cite{white}. We shall present
results for open spin chains with spin values $S=1/2,1,3/2,2$ up to chain
length of 100 sites.

  The DMRG method has proved to be tremendously successful in studying $S=1$
and $S=1/2$ antiferromagnetic Heisenberg spin chains\cite{white,wh}. The method
was found to be particularly suitable for studying spin chains with open
boundary condition and is thus well suited for our purpose of studying edge
states. We use the infinite chain algorithm\cite{white} in our study. Two
new sites are added to the spin chain in each optimal step of the calculation
from length $L=4$ to $L=100$. That is, open spin chains with $even$ number
of sites are studied. The number of the kept optimized states in our
calculation is $m=120$. The largest truncation errors for $S=1/2,1,3/2$, and
$2$ chains are found to be smaller than $10^{-10}$, $10^{-8}$,
$3\times 10^{-6}$ and $10^{-5}$ respectively
when the chain length reaches $L=100$ in the final step.
Properties of the ground state and a few lowest excited
states are obtained by looking at the lowest energy state
with fixed total z-component of spin angular momentum $S_z^{tot}$.
In particular, the ground state corresponds to the lowest energy state
in the sector $S_z^{tot}=0$. We shall look at the excitation energies
of various states with different $S_z^{tot}$ and the corresponding
average z-component of angular momentum on each site i
$= <S_z^{i}>$. The dimmerization parameter $q(i)=<S_{i}.S_{i+1}-S_{i-1}.S_{i}>$
for the ground state will also be examined.

  We start with the excitation energies for integer spin chains. Fig.1 shows
the excitation energies $E_{n}-E_{0}$ for $n=1$ to $3$ as a function of chain
length $L$ for the $S=1$ spin chain. $E_{n}$ is the energy of the
lowest energy state in the sector $S_z^{tot}=n$. Energy is measured in units
of Heisenberg coupling J. According to the valence bond
picture\cite{aklt,ng2}, for $S=1$ spin chain, two $S=1/2$ spins are left at
two ends of the spin chain, and are coupled with effective coupling $J_{eff}
\sim Je^{-L/\xi}$, where $\xi$ is the correlation length. For even spin
chains, the coupling is antiferromagnetic and the resulting ground state is a
spin singlet($S_z^{tot}=0$). The lowest energy state with
$S_z^{tot}=1$ can be constructed by exciting the singlet formed by the two
edge spins into a triplet, and the excitation energy is of order $J_{eff}$,
which goes to zero exponentially as length of spin chain increases. Excited
states with larger $S_z^{tot}$ cannot be constructed by exciting only the
edge spins any more and bulk excitations must be involved in constructing
states with $S_z^{tot}>1$, implying that the excitation energy will be of
order $E_{n}-E_{0}\sim (n-1)\times{E_H}$ as $L\rightarrow\infty$,
where $E_H$ is the Haldane gap. This predicted behavior is clearly confirmed
by our numerical result presented in Fig.1. The Haldane gap $E_H$ and
correlation length $\xi$ are estimated to be $E_H={0.41J}$ and
$\xi ={6.0}$
from $E_2-E_0 \to 0.41 J$ and $E_1-E_0 \sim 0.7 e^{-L/6.0}$ for large $L$
respectively, in agreement with result obtained by
White\cite{white}. Fig.2 shows similar results for $S=2$ spin chain. Notice
that in this case, the $S_z^{tot}=2$ state also has excitation energy going
to zero as $L\rightarrow\infty$, whereas higher $S_z^{tot}$ states have finite
energy gap.
This behavior is consistent with valence bond picture which predicts
that for $S=2$ spin chain, the edge state spin magnitude is $S/2=1$, and the
two edge spins are coupled again by $J_{eff}\sim{Je^{-L/\xi}}$, where $\xi$ is
now the correlation length for $S=2$ spin chain.
In this case both $S_z^{tot}=1$ and $S_z^{tot}=2$ states can be constructed
by exciting edge spins only, and have excitation energy $\rightarrow{0}$
as $L\rightarrow\infty$, which is exactly the
behavior obtained in our numerical calculation. The Haldane gap and
the correlation length are estimated to be $E_{H}\sim 0.02$ and
$\xi\sim 33$ from $E_3-E_0 \to 0.02 J$, $E_1-E_0 \sim 0.1 e^{-L/33}$,
and $E_2-E_0 \sim 0.4 e^{-L/33}$ for large $L$
, respectively from our numerical results. However, the accuracy
of these estimated numbers are much worse than the $S=1$ case
because of the much larger truncation error found in our
calculation for the $S=2$ spin chain.

Next we turn to the excitation energies for half-integer spin chains. Fig.3
shows the excitation energies $E_{n}-E_{0}$ for $n=1$ to $3$ as
a function of inverse chain length $L^{-1}$
for $S=1/2$ spin chain. According to (abelian) bosonization theory\cite{aff}
the excitation energy is given by $E_{n}-E_{0}=(S_z^{tot})^2\times(\pi/L)$.
The bosonization prediction is also drawn on
Fig.3 (dash line). Notice that bosonization theory is correct only in the
asymptotic limit $L\rightarrow\infty$ and terms of order $(Lln(L))^{-1}$
are  neglected\cite{aff}. It is thus not surprising that our numerical
results and bosonization theory predictions show only qualitative
agreement with each other. Notice also that bosonization theory
for finite $S=1/2$ spin chain predicts no edge states in this case.
Fig.4 shows similar results for $S=3/2$ spin chain where the excitation
energies are shown up to state with $S_z^{tot}=4$.
It is apparent that $E_{n}-E_{0}\sim{1/L}$ in all four cases.
Notice however, that edge state
with spin magnitude $S=1/2$ is predicted to exist in this case, according
to the conjecture by Ng\cite{ng}. However it does not seem to show
up clearly as for integer spin chains in the energy curve.

To understand this behavior we examine the edge state in $S=3/2$ spin chain
in more detail. According to Ng\cite{ng}, the low energy physics of open
$S=3/2$ spin chain can be described by an effective $S=1/2$ spin chain
coupled antiferromagnetically to two impurity spins with magnitude
$S_{im}=1$ at two ends of the spin chain. The impurity spin will be partially
screened by a Kondo type effect at low energy\cite{ea}, leaving ``free''
spins of magnitude $S_{edge}=1/2$ at the ends, coupling $ferromagnetically$
to the bulk $S=1/2$ spin chain. Again, the coupling of the ``impurity'' spin
to the bulk $S=1/2$ spin chain can be analyzed using renormalization group
technique. For ferromagnetic coupling, the corresponding operator is
marginally irrelevant, implying that a free $S_{edge}=1/2$ is left at each end
of the spin chain as $L\rightarrow\infty$. For a finite spin chain, an RKKY
type coupling between the two edge spins will be found, and the resulting
ground state is a spin-singlet for even chains, as is in the case of integer
spin chains. The RKKY coupling $J_{R}$ between the two edge spins comes from
exchange of spinwaves, and has a length dependence $J_{R}\sim{g/L}$,
where $g$ is some effective coupling constant. Thus we expect that the
excitation energy for the edge spins $E_{ed}$ is proportional to $g/L$.
Notice that at large distance $L$, the
coupling constant $g$ will be renormalized with
$g\rightarrow{g/(1+gln(L))}$ (ferromagnetic Kondo effect) and
corresponding correction to energy $E_{ed}$ will be found. In
particular, at large $L$, $E_{ed}\sim 1/(Lln(L))$.
Assuming that the low energy bulk excitations are described by
bosonization theory as an effective $S=1/2$ spin chain, we conclude
that the lowest ($S_z^{tot}=1$) excited state for $S=3/2$ open spin chain
is an edge excitation as in integer spin chains, since the bulk excitation
energies
scale as $1/L$, and will be always higher in energy then $E_{ed}$ as
$L\rightarrow\infty$. More detailed analysis of the energy spectrum supports
our
edge state energy analysis which we shall discuss in a later paper. In the
following, we shall present a more direct numerical evidence supporting our
edge states picture for Heisenberg spin chains.

  Fig.5 shows the expectation value $<S_z^{i}>$ for the $S_z^{tot}=1$ state
for all four spin chains $S=1/2,1,3/2,2$ with 100 sites, i.e. $i=1,100$.
The four cases are arranged from top to bottom in increasing order
of spin magnitude, i.e. the top one is for $S=1/2$, and bottom one for $S=2$,
etc. For spin singlet ground states, $<S_z^{i}>$ will be zero for all
sites $i$. Thus $<S_z^{i}>$ for the $S_z^{tot}=1$ state gives information about
the wavefunction of the first excited state in the spin system. Notice that for
all four spin values, Fig.5 shows that the excited states all carry staggered
magnetization because of the underlying antiferromagnetic
interaction. For $S=1/2$ spin chain, Fig.5
suggests clearly that the first excited state can be thought of as
a standing spin wave. The same qualitative result is also obtained from
bosonization theory of finite $S=1/2$ spin chain. In fact the feature can be
understood in the much simpler $S=1/2$ XY-model with $J_z=0$. The $S=1/2$
XY-model can be mapped onto a non-interacting spinless fermion model. It is
easy to show that similar feature exists in this case. The introduction of
nonzero $J_z$ term just modified the feature quantitatively. It is clear
from Fig.5 that the wavefunction of the first excited states in the $S>1/2$
spin chains are qualitatively different from that of the $S=1/2$ spin chain.
In fact, in all three cases we considered, $<S_z^{i}>$ suggests clearly that
the first excited states are all consist of spin excitations which are
localized
around edges of spin chains, i.e. they are edge excitations. It is also clear
from the figure that the correlation length of the edge state in $S=2$
spin chain is much larger than the corresponding correlation length of the
$S=1$ spin chain, in agreement with general expectation and with results
obtained from excitation energy analysis. One can also study
$<S_z^{i}>$ for the $S_z^{tot}=2$ states and looks
at the difference in $<S_z^{i}>$ between the $S_z^{tot}=2$ and $S_z^{tot}=1$
states. For the $S=3/2$ spin chain, the difference is found to have a
spinwave like feature similar to $<S_z^{i}>$ for $S=1/2$ spin chain with
$S_z^{tot}=1$, suggesting clearly that the spin magnitude of the
edge state is $S_{edge}=1/2$, and excitations with $S_z^{tot}>1$
involves bulk excitations, in agreement with previous conjecture by
Ng\cite{ng}.

   We have also examined the dimmerization parameter $q(i)=<S_{i}.S_{i+1}-
S_{i}.S_{i-1}>$ for the ground states of the four spin chains we considered.
For integer spin chains, $q(i)$'s are found to be nonzero around ends of
spin chains and decay exponentially as one move into the interior, in
agreement with prediction by Ng\cite{ng,ng2}. For half-integer spin chains,
$q(i)$'s are found to be large at the ends and decay with power law
into the interior of spin chain. Similar results
were also obtained by White\cite{white} for $S=1$
and $S=1/2$ chains. For $S=1/2$ chain, the dimmerization parameter $q(i)$
can again be examined using bosonization techniques where it can be shown that
$q(L/2)\sim (1/L)^{\beta}$, where $\beta=1/2$ in bosonization theory. Similar
behavior is observed in our numerical results for both $S=1/2$ and $S=3/2$
chains, with $\beta \sim 1/2$ for $S=1/2$ chain and $\beta \sim
0.6$ for $S=3/2$ chain. The huge difference in behaviour of $q(i)$
between integer and half-integer open spin chains reflects the fact
that half-integer spin chains are more susceptible to spin-Peierls
instability and furnish another interesting parameter
distinguishing between integer and half-integer spin chains.

  Lastly, we want to list our calculated ground state energies (per site)
$\epsilon_0$ for the four spin chains. We obtain
$\epsilon_0=-0.443147$ for $S=1/2$ chain and $\epsilon_0=-1.401484$
for $S=1$ chain, in good agreement with exact Bethe
ansatz solution and the result obtained by S.White et.al \cite{wh}
respectively. We have also computed the ground state
energies for $S=3/2$ and $S=2$ chains with good precision, with
$\epsilon_0=-2.8283$ for $S=3/2$ chain and $\epsilon_0=-4.7608$
for $S=2$ chain.

  Summarizing, in this paper we have performed numerical studies on low energy
properties of finite quantum spin chains with spin values $S=1/2,1,3/2,2$ using
the Density-Matrix Renormalization Group techniques which has been applied
successfully to study $S=1/2$ and $S=1$ spin chains. We find that all the
results we obtained are consistent with the edge states picture conjectured by
Ng\cite{ng}. In particular, we present for the first time numerical evidence
for
existence of edge states in $S=3/2$ and $S=2$ finite quantum spin chains. The
dimmerization parameter $q(i)$'s for the ground states are also studied where
the huge difference in behavior between integer and half-integer spin chains
is pointed out.

S. Qin and Z.B. Su would like to thank Prof. C.L. Wang,
Prof. Lu Yu for their various helps and suggestive discussions.
One of the authors (ZBS) thanks Prof. Clare Yu and Prof. Shoudan Liang
for interesting discussions, he would
acknowledges the warm hospitality extended to him
by the Physics Department of HKUST last spring. This work is
partially supported by NSFC, CCAST, LSEC and UPGC Grant UST636/94P.


\begin{figure}
\caption{
The excitation energy $E_n-E_0$ for even $S=1$ chain as function of
chain length $L$.
The black circles, black squares, and black triangles are for
    $E_1-E_0$,     $E_2-E_0$    , and $E_3-E_0$ respectively.
}
\end{figure}

\begin{figure}
\caption{
The excitation energy $E_n-E_0$ for even $S=2$ chain as function of
chain length $L$.
The open circles, black circles, black squares, and black triangles are for
    $E_1-E_0$, $E_2-E_0$, $E_3-E_0$, and $E_4-E_0$ respectively.
}
\end{figure}

\begin{figure}
\caption{
The excitation energy $E_n-E_0$ for even $S=1/2$ chain as function
of inverse chain length $1/L$.
The black circles, black squares, and black triangles are for
    $E_1-E_0$, $E_2-E_0$, and $E_3-E_0$ respectively.
The dashed lines are $E_n-E_0=n^2\; \pi /L$.
}
\end{figure}

\begin{figure}
\caption{
The excitation energy $E_n-E_0$ for even $S=3/2$ chain as function
of inverse chain length $1/L$.
The open circles, black circles, black squares, and black triangles are for
    $E_1-E_0$, $E_2-E_0$, $E_3-E_0$, and $E_4-E_0$ respectively.
}
\end{figure}

\begin{figure}
\caption{
The configuration $<S_z^i>$ for the lowest states
of $S=1/2,1,3/2,2$ chains with $S_z^{tot}=1$ and length $L=100$.
The zero point for $S=1/2,1,3/2,2$ chains are shifted to $4,3,2,1$
respectively.
The circles, black circles, black squares, and black triangles are for
the $<S_z^i>$ of chains with $S=1/2,1,3/2,2$ respectively.
}
\end{figure}

\end{document}